\documentclass[twocolumn,secnumarabic,amssymb,superscriptaddress, nobibnotes, nofootinbib,aps, prd]{revtex4-1}

\usepackage{mathrsfs}
\usepackage{physics}
\usepackage{tikz}
\usepackage[caption=false]{subfig}
\usepackage[colorlinks=true,linkcolor=blue,citecolor=blue,urlcolor=blue]{hyperref}
\usepackage{cleveref}
\usepackage{amsmath,amssymb,amsfonts,amsthm}
\usepackage{graphicx}
\usepackage{ragged2e}
\usepackage{float}
\usepackage{bm}

\def\rar{\rightarrow}

\def\p{\partial}
\def\ra{\rangle}
\def\la{\langle}

\def\no{\nonumber}
\def\bea{\begin{eqnarray}}
\def\eea{\end{eqnarray}}
\def\be{\begin{equation}}
\def\ee{\end{equation}}

\def\p{\partial}
\usepackage{filecontents}

\begin{document}

\title{Relative entropy formulation of thermalization process in a Schwarzschild spacetime}%
\author{Si-Wei Han}
\affiliation{School of Physics, Xi'an Jiaotong University, Xi'an, Shaanxi 710049, China}

\author{Zhichun Ouyang}
\affiliation{School of Physics, Xi'an Jiaotong University, Xi'an, Shaanxi 710049, China}

\author{Zhiyao Hu}
\affiliation{School of Physics, Xi'an Jiaotong University, Xi'an, Shaanxi 710049, China}

\author{Jun Feng}%
\email{Corresponding author:\\
j.feng@xjtu.edu.cn}
\affiliation{School of Physics, Xi'an Jiaotong University, Xi'an, Shaanxi 710049, China}
\affiliation{Institute of Theoretical Physics, Xi'an Jiaotong University, Xi'an, Shaanxi 710049, China}

\date{\today}%

\begin{abstract}

We revisit the problem of the thermalization process in an entropic formulation for the Unruh-DeWitt (UDW) detector outside a Schwarzschild black hole. We derive the late-time dynamics of the detector in the context of open quantum system, and capture the path distinguishability and thermodynamic irreversibility of detector thermalization {process} by using quantum relative entropy (QRE). We find that beyond the Planckian transition rate, the refined thermalization process in detector Hilbert space can be distinguished by the time behavior of the related QRE. We show that the exotic position-dependent behaviors of the QRE emerge corresponding to different choices of black hole vacua (i.e., the Boulware, Hartle-Hawking, and Unruh vacua). Finally, from a perspective of quantum thermodynamics, we recast the free energy change of the UDW detector undergoing Hawking radiation into an entropic combination form, where the classical Kullback-Leibler divergence and quantum coherence are presented in specific QRE-like forms. {With growing Hawking temperature, we} find that the consumption rate of quantum coherence is larger than that of its classical counterpart.
\end{abstract}


\maketitle


\section{Introduction}
\label{1}

The fascinating side of a black hole is its unique geometry, which can amplify short-distance fluctuations to macroscopic sizes. In his series of seminal papers around the 1970s', Hawking predicted \cite{Hawking1,Hawking2} that such stretching of fluctuations causes radiation from the event horizon and makes a black hole evaporate. However, this process is inconsistent with the unitarity of quantum mechanics, i.e., pure states can only evolve to other pure states, leading to the black hole information paradox \cite{BH1}, which has instigated a large amount of research in several decades. 

The resolution of the paradox requires a proper understanding of the thermal nature of Hawking radiation. Utilizing local quantum probes, e.g., a particle detector model referred to as an Unruh-DeWitt (UDW) detector \cite{Hawking3,UDW1}, the Hawking effect can be demonstrated by the thermalization \emph{end} of the detector state. That is, outside a black hole, the detector coupling to a background field would be excited and asymptotically driven to a Gibbs state with effective temperature $T_{\text{eff}}$, which approaches the Hawking temperature $T_H$ in the spatial infinity, regardless of its initial state \cite{UDW2}.

In recent years, from an alternative perspective of the so-called open quantum system framework \cite{Open1}, the UDW detector is viewed as a local open system, while background quantum fluctuations play the role of environment that induces dissipation and decoherence \cite{Open2}. The method is particularly powerful when one wishes to understand long-time processes such as thermalization \emph{process} of a simple quantum system due to its late-time dynamics, which can be fully achieved now. For example, it allows one to directly determine if and how a UDW detector undergoing Hawking-Unruh effect approaches a Gibbs state \cite{Open3,Open4,Open5,Open6}, rather than stopping at the detailed balance condition or Planckian transition rates \cite{Open7,Open8}, which can not capture detector density matrix decoherence, thus only be \emph{necessary but not sufficient} conditions for thermalization. 

Once knowing the dynamics of a detector, the pressing problem is naturally to find some \emph{feature functions} to characterize the time-evolution details of an open process. Beyond response functions in the "conventional" field approach \cite{Open13}, it was proposed to be quantum Fisher information (QFI) that resolves the statistics inversion in Unruh effect \cite{Open14} and metrologically probes particular spacetime structures \cite{Open15,Open16,Open17}, or entropic uncertainty bound (EUB) that witnesses the information loss from Hawking-Unruh decoherence \cite{Open18,Open19}. Very recently, quantum coherence monotone, an ultimate feature setting quantum and classical worlds apart \cite{Coherence1}, has been suggested as a quantum resource part during the thermalization process of an accelerating UDW detector \cite{Open20,Open21} with anyonic response spectrum \cite{Open22}. 

In this work, we revisit the thermalization process of a UDW detector outside a Schwarzschild black hole, appealing to a {refined} feature function known as \emph{quantum relative entropy} (QRE). Between quantum states $\rho$ and $\sigma$, it is defined as
\be
S(\rho \| \sigma)\equiv\operatorname{tr}(\rho \log \rho)-\operatorname{tr}(\rho \log \sigma)   \label{eq1.1}
\ee
which is always positive and becomes vanishing only if $\rho=\sigma$. Our interests in QRE are two-fold:

(1) Intuitively, the irreversible thermalization of the detector corresponds to a continuous evolution of its density matrix in Hilbert space towards a thermal end $\sigma_{\text{th}}$. We thus expect QRE, the most appropriate distance-type measure among various tools in quantum information \cite{Coherence2}, to characterize \emph{state distinguishability} between the late-time state $\rho_{\text{detector}}(t)$ of the UDW detector and equilibrium state $\sigma_{\text{th}}$ at Hawking temperature
\be
\mathcal{D}\equiv S(\rho_{\text{detector}}(t) \| \sigma_{\text{th}})   \label{eq1.2}
\ee
whose time-evolution encodes the thermalization {process} in the detector's state space. 

(2) The QRE plays a profound role in a quantum thermodynamic description of irreversible open processes (see Ref. \cite{qthermo1} for a review). For a general non-equilibrium density matrix, its free energy $F(\rho)=\operatorname{tr}(\rho H)-T S(\rho)$, {where $S({\rho})=-\operatorname{Tr}({\rho} \log {\rho})$ is von Neumann entropy}, can be measured with respect to an equilibrium reference state as $F(\rho)=F(\sigma_{\text{th}})+TS(\rho \| \sigma_{\text{th}})$. An important insight is that with a specific QRE, non-equilibrium free energy can further be separated into a form with genuine classical and quantum contributions \cite{qthermo3}. Falling into a UDW detector regime, this indicates that the non-equilibrium free energy undergoing open dynamics can similarly be recast into an entropic formulation \cite{qthermo4}
\be
F\left(\rho_{\text{detector}}\right)=F(\sigma_{\text{th}})+T\mathcal{D}_{KL}+ T\mathcal{C}   \label{eq1.3}
\ee
Here, $\mathcal{D}_{KL}$ is Kullback-Leibler divergence of the populations of density matrix \cite{Coherence3}, a purely classical term, and $ \mathcal{C}$ is quantum coherence quantified in terms of QRE as \cite{Coherence1}
\be
\mathcal{C}\equiv S(\rho_{\text{detector}}(t)\| \rho_{\text{diag}})   \label{eq1.4}
\ee
where $\rho_{\text{diag}}$ is an incoherent state derived from the dephasing of the detector's state.

Motivated by the above considerations, the main content of the work is to explore the thermalization process of a UDW detector in a Schwarzschild spacetime from two perspectives, i.e., (1) distinguishability between different {ways that the detector approaches its thermalization end} (2) the genuine quantum contribution driving thermodynamic irreversibility, by using two specific forms of QRE respectively.

We first give the late-time dynamics of the detector in an open quantum system framework, with respect to delicately vacuum choices (i.e., Boulware, Hartle-Hawking, and Unruh states corresponding to different timelike Killing vectors) of the background quantum field. In particular, in the Hartle-Hawking vacuum defined by the positive modes with respect to horizon generators in the Kruskal extension, the UDW detector will be excited into an equilibrium end with the Planckian spectrum, indicating the well-known Hawking effect.

To characterize a specific thermalization {process} of the UDW detector, we compute the associated QRE (\ref{eq1.2}). We are interested in how the local properties of the detector (e.g., initial state preparation) and vacua choice for black hole may affect the time evolution of QRE, corresponding to different thermalization trajectories of the detector but with the same asymptotic thermal end. 

To clarify the genuine quantum nature of detector thermalization, we compute the quantum coherence quantifier (\ref{eq1.4}) in non-equilibrium free energy. The calculation based on a quantum thermodynamic perspective just complements previous studies \cite{Open20,Open21,Open22}. Since the irreversible open dynamics of the detector are significantly influenced by its perceived Hawking radiation, the monotone (\ref{eq1.4}) also unveils how the black hole quantum effect can determine the non-equilibrium thermodynamics stemming from (\ref{eq1.3}).

The paper is organized as follows. In Section 2, we solve the late-time dynamics of a UDW detector outside the Schwarzschild black hole. In Section 3, the QRE (\ref{eq1.2}) is calculated for the detector interacting with field fluctuations in Boulware, Hartle-Hawking, and Unruh vacuum, respectively.  In Section 4, the time-evolution of quantum coherence (\ref{eq1.4}) is studied, which determines the non-equilibrium thermodynamics of the open process outside the black hole. In Section 5, the summary and discussion are given. For simplicity, throughout the analysis, we use natural units $\hbar=c=k_{B}=1$.

\section{Open dynamics of the detector outside a black hole}
\label{2}
For the sake of completeness, we begin with a brief review of the open quantum system approach to the standard Hawking effect, following closely to that in Ref.\cite{Open3}. We consider a UDW detector hovering outside a Schwarzschild black hole whose geometry is described as 
\be
d s^{2}=\left(1-\frac{2 M}{r}\right) d t^{2}-\left(1-\frac{2 M}{r}\right)^{-1}d r^{2}-r^{2} d \bm{\Omega}^{2}     \label{eq2.1}
\ee
The detector interacts with a massless scalar field, which can be regarded as an environment.

\subsection{Master equation for single UDW detector}
The total Hamiltonian of the combined system of detector and quantum field is
\be
H=H_{\text{detector}}+H_{\Phi}+g H_{\text{int}}              \label{eq2.2}
\ee
where UDW detector is modeled by a two-level atom $H_{\text{detector}}=\frac{1}{2}\omega \sigma_3$, $H_\Phi$ is the Hamiltonian of free scalar field $\Phi(x)$, which satisfies the standard Klein-Gordon equation in Schwarzschild spacetime. The atom-field interaction is given by $H_I=(\sigma_++\sigma_-)\Phi(x)$, with the atomic raising and lowering operators as $\sigma_\pm$, and $\omega$ is the energy level spacing of the atom. 

The time evolution of the combined system is driven by the Hamiltonian (\ref{eq2.2}), according to von Neumann equation
\be
\frac{d{\rho}_{tot}}{d\tau}=-i[H,\rho_{tot}(\tau)]             \label{eq2.3}
\ee
where $\tau$ is the proper time of the detector.

To achieve the reduced dynamics of the detector, three typical assumptions need to be introduced. Firstly, one assumes a weak coupling between the detector and the external field, and writes the combined state in a factorized form (\emph{Born approximation}), e.g., $\rho_{tot}(0)=\rho(0)\otimes |0\ra\la0|$, where $\rho(0)$ is the detector initial state and $|0\ra$ is the field vacuum in the comoving frame. This approximation is justified when the environment is large enough compared to the system, ensuring that the von Neumann equation can be perturbed. By integrating over the background field degrees of $\rho_{tot}(\tau)$, the dynamics of the UDW detector are reduced to an integro-differential master equation
\be
\frac{d \rho}{d \tau}=-i [H_{\text{detector}},\rho(\tau)]+\int_{0}^{\tau} d s~ \mathcal{K}_s[\rho(\tau-s)]          \label{eq2.4}
\ee
In general, the integral contains kernel $\mathcal{K}_s$, which is a series expansion in $H_{\text{int}}$ is very complicated and depends on the entire history of detector evolution, thus making the master equation be intractable. To simplify the problem, one employs the \emph{Markov approximation}, which neglects the memory effect of the environment once the typical variation time of the detector is larger compared to the decay time of the field correlation. This enables the integration limit in (\ref{eq2.4}) to be pushed to infinity \cite{Open1}. Finally, a \emph{secular approximation} \footnote{The bounds for the validity of these approximations should be taken with subtle care. For example, the performed Markovian limits are allowed in general for the late-time open dynamics while in early-time non-Markovian effect should be included \cite{Open4}. Also, the secular approximation admits a more narrow parameter space than usually dose \cite{bound1}. Nevertheless, for \emph{single} detector \cite{bound2}, all analysis for the late-time dynamics is still reliable under these cares. } is used to average all rapidly oscillating terms in the master equation, eventually making the detector's dynamical evolution follow a completely positive and trace-preserving (CPTP) map $\rho(0)\mapsto\rho(\tau)$. The reduced detector dynamics is then governed by the Gorini-Kossakowski-Sudarshan-Lindblad (GKSL) master equation \cite{Open23,Open24}
\be
\frac{d\rho}{d\tau}=-i\left[H_{\text{eff}},\rho(\tau)\right]+\sum^3_{i,j=1}C_{ij}\left[\sigma_j\rho \sigma_i-\frac{1}{2}\left\{\sigma_i\sigma_j,\rho\right\}\right]    
\label{eq2.5}
\ee

Compared to the von Neumann equation (\ref{eq2.3}), two modifications in (\ref{eq2.5}) arise from the interaction between the detector and external fields. Firstly, a dissipation emerges as the second term on the right-hand side of (\ref{eq2.5}), where the elements of the Kossakowski matrix are
\be
C_{ij}=\frac{\gamma_+}{2}\delta_{ij}-i\frac{\gamma_-}{2}\epsilon_{ijk} n_k+\gamma_0\delta_{3,i}\delta_{3,j}     \label{eq2.5+}
\ee
with coefficients
\be
\gamma_\pm= \mathcal{G}(\omega)\pm \mathcal{G}(-\omega),~~~~\gamma_0=\mathcal{G}(0)-\gamma_+/2         \label{eq2.6}
\ee
defined in terms of two-side Fourier transformation on the Wightman function of the background field
\be
\mathcal{G}(\omega)=\int_{-\infty}^{\infty}d\tau~e^{i\omega \tau}\langle0|\Phi[x(\tau)] \Phi[x(0)]|0\rangle
  \label{eq2.6+}
\ee

{Besides producing the nonunitary evolution, the interaction to the external scalar field $\Phi$ induces so-called Lamb shift correction $H_L$ to the detector Hamiltonian, like $H_{\text {eff}}=H_{\text{detector}}+H_L$. As for the Kossakowski matrix, this additional shift $H_L$ can be expressed in terms of the field correlations \cite{Open3}, leading to} $H_{\text{eff}}=\frac{1}{2}{\Omega}\sigma_3$, where the energy levels of the detector are undergone a Lamb-type renormalization, i.e., ${\Omega}=\omega+i[\mathcal{K}(-\omega)-\mathcal{K}(\omega)]$, where $\mathcal{K}(\lambda)=\frac{1}{i\pi}\text{P}\int_{-\infty}^{\infty}d\omega\frac{\mathcal{G}(\omega)}{\omega-\lambda}$ with $\text{P}$ denoting the principal value.

\subsection{The late-time dynamics and its thermalization end}

For the two-level UDW detector, its density matrix can be expressed in a Bloch form
\be
\rho(\tau)=\frac{1}{2}\Big(1+\sum_{i=1}^3n_i(\tau)\sigma_i\Big)               \label{eq2.7}
\ee 
where the Bloch vector $\bm{n}(\tau)$ has its length $\ell=\sqrt{\sum_{i=1}^3n_i^2}$ to characterize its purity.

Substituting it into the master equation (\ref{eq2.5}), with assuming a general initial state $|\psi\ra=\sin\frac{\theta}{2}|0\ra+\cos\frac{\theta}{2}|1\ra$, the density matrix (\ref{eq2.7}) can be explicitly resolved in terms of time-dependent elements
\bea
n_1(\tau)&=&\mathcal{E}(\tau)\sin\theta\cos{\Omega}\tau,\no\\
n_2(\tau)&=&\mathcal{E}(\tau)\sin\theta\sin{\Omega}\tau,\no\\
n_3(\tau)&=&\mathcal{E}^2(\tau)\left(\cos\theta+\gamma\right)-\gamma            \label{eq2.8}
\eea
where the decay rate $\mathcal{E}(\tau)\equiv \exp(-\gamma_+\tau/4)$ and $\gamma\equiv\gamma_-/\gamma_+$ are introduced. The length of the Bloch vector $\bm{n}(\tau)$ at arbitrary time is
\be
\ell(\tau)=\sqrt{\left[\mathcal{E}^2(\cos\theta+\gamma)-\gamma\right]^2+\mathcal{E}^2\sin^2\theta}         \label{bloch1}
\ee

Due to the atom-field interaction, the detector state evolves nonunitarily to a mixed equilibrium. It is easy to observe from (\ref{eq2.8}) that in the limit $\tau\rightarrow\infty$ the asymptotic state is
\be
\sigma_{\text{th}}=\frac{1}{2}\left(\begin{array}{cc}1-\gamma & 0 \\0 & 1+\gamma\end{array}\right)\equiv\frac{e^{- H_{\text{detector}}/T_\text{eff}}}{\operatorname{Tr}\left[e^{- H_{\text{detector}}/{T}_\text{eff}}\right]}             \label{eq2.9}
\ee
which has a Gibbs form with an effective temperature
\be
T_{\text{eff}}=\frac{\omega}{2\tanh^{-1}(\gamma)}        \label{eq2.10}
\ee
This indicates a unique thermalization end for the detector undergoing open dynamics.

For the UDW detector hovering outside a Schwarzschild black hole, the thermal nature of Hawking radiation can be unveiled by computing the rate of state transition of the detector \cite{Open13}, i.e., $\mathcal{P}_{i \rightarrow f}(\tau)=\operatorname{Tr}\left[\rho_{f} \rho(\tau)\right]$, between arbitrary initial and final states of the detector. Once choosing initially $\rho(0)\equiv\rho_{i}$ and $\rho_{f}$ are the ground and excited states of the detector, the probability for a spontaneous transition of the detector is \cite{Open2}
\be
\mathcal{P}_{\uparrow}=\frac{\gamma_+-\gamma_-}{2 \gamma_+}\left(1-e^{-2 \gamma_+ \tau}\right).     \label{eq2.11}
\ee
which gives the transition probability per unit time at $\tau=0$ as
\be
\Gamma_{\uparrow}=\left.\frac{\partial \mathcal{P}_{\uparrow}}{\partial \tau} \right|_{\tau=0}=\mathcal{G}\left(-\omega\right)    \label{eq2.12}
\ee
In a curved spacetime, it is delicate to choose the proper vacuum states of the background field, associated with some timelike Killing vectors. In the so-called Hartle-Hawking vacuum defined by the positive modes with respect to past (future) horizon generators $\p_U$ ($\p_V$) in Kruskal extension, the transition rate has a Planckian form \cite{Open3}
\be
\Gamma_{\uparrow,H}\propto\frac{\omega}{\pi\left(e^{2 \pi \omega / \kappa_r}-1\right)}
\label{eq2.13}
\ee
where $\kappa_{r}=\kappa/\sqrt{g_{00}}$ and $\kappa=1/4M$ is the surface gravity of Schwarzschild black hole. Eq.(\ref{eq2.13}) implies what is usually referred to as the Hawking effect. 

\begin{figure}[hbtp]
\begin{center}
\includegraphics[width=.3\textwidth]{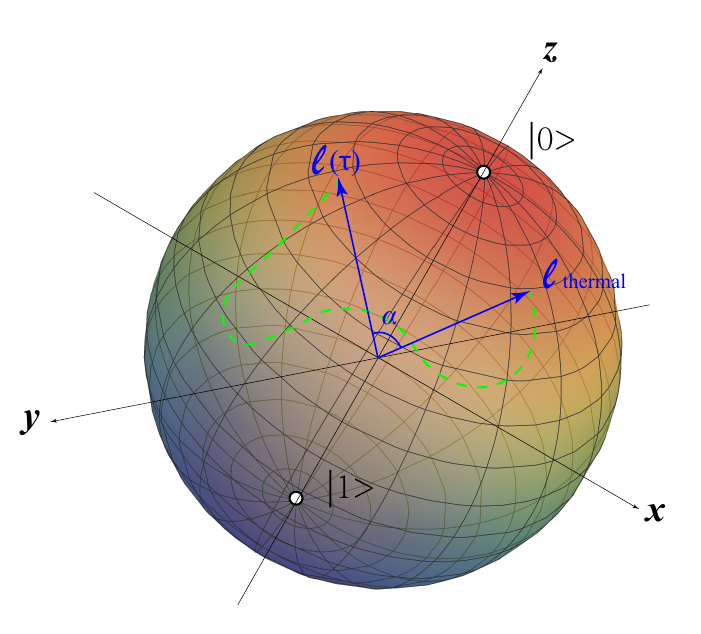}
\caption{The detector state is uniquely labeled by the vector $\bm{n}$ on a Bloch sphere, with length $\ell\leq1$. The intersection points of the axes represent the eigenstates of Pauli operators, e.g., $|0\ra$ and $|1\ra$ for $\sigma_3$. The green dashed curve {encodes} a continuous thermalization {process} of the detector towards an equilibrium end with $\bm{n}_{\text{thermal}}$.}
\label{fig1}
\end{center}
\end{figure}

Nevertheless, the emerging Planckian transition rate is just necessary but not sufficient conditions for thermalization. What is truly indicated by  (\ref{eq2.13}) is a (classical) rearrangement of populations but neglected off-diagonal (quantum) coherences of detector density matrix \cite{qthermo3}. Once accounting for both classical and quantum sides of the process, even for the same end of thermalization, the detector may follow different paths in its Hilbert space as mapped onto a Bloch sphere (see Fig.\ref{fig1}). In this way, the late-time dynamics as well as the thermalization process of the UDW detector can be identified as a continuous trajectory in the Bloch sphere, following which the detector Bloch vector $\bm{n}(\tau)$ evolves to a thermal state with length $\ell_{\text{thermal}}$. As a proper distance measure along the trajectory, we expect the time evolution of QRE to encode complete information of any specific thermalization path.


\section{QRE of the detector in black hole spacetime}
\label{sec3}

We use the QRE (\ref{eq1.2}) as a probe to measure the distance between the UDW detector state $\rho(\tau)$ and a Gibbs state (\ref{eq2.9}) representing a thermalization end after a sufficiently long time. As QRE depends on open dynamics of density matrix $\rho(\tau)$, the quantity $\mathcal{D}(\tau)$ can distinguish various thermalization trajectories, encoding the information of detector's local features (e.g., initial state preparation) or black hole geometry. 

Using Cortese formula \cite{Coherence3}, we calculate the QRE for single-qubit states $\rho$ and $\sigma_{\text{thermal}}$ as
\bea
\hspace*{-20pt}\mathcal{D}(\tau)
&=&\operatorname{tr}(\rho \log \rho)-\operatorname{tr}(\rho \log \sigma_{\text{thermal}}) \no\\
&=&\frac{1}{2}\log\left(\frac{1-\ell^2}{1-\ell_{\text{thermal}}^2}\right)+\frac{\ell}{2}\log\left(\frac{1+\ell}{1-\ell}\right)\no\\
&&~~~~~~~~~~~ -\frac{\ell_{\text{thermal}}\cos\alpha}{2}\log\left(\frac{1+\ell_{\text{thermal}}}{1-\ell_{\text{thermal}}}\right)                             \label{eq3.1}
\eea
where $\cos\alpha=-n_3/\ell$ is the angle between the two Bloch vectors (see Fig.\ref{fig1}) and $\ell_{\text{thermal}}=\gamma$ as observed from (\ref{eq2.9}).

Like aforementioned, in a black hole spacetime, quantum fields are complicated by a delicately choice of proper vacuum states with respect to particular timelike Killing vectors. In the spacetime of a Schwarzschild black hole, we have three kinds of proper vacuum states, i.e., the Boulware vacuum, the Hartle-Hawking vacuum, and the Unruh vacuum. In the following, we calculate the QRE (\ref{eq3.1}) for three vacua respectively. For numerical estimation, we introduce three dimensionless parameters 
\be
\tilde{\tau}\equiv\omega\tau,~~~~~\tilde{R}\equiv\frac{r}{r_H},~~~~~ \tilde{T}_H\equiv {T_H}/{\omega}=\frac{\kappa}{2\pi\omega}             \label{eq3.8}
\ee
which are the rescaled proper time, the relative distance of the detector measured in Schwarzschild radii $r_H=2M$, and the rescaled Hawking temperature. 
 

 \subsection{The Boulware vacuum}
The Boulware vacuum corresponds to the familiar concept of an empty state for large radii, defined by the positive frequency modes with respect to a Schwarzschild timelike Killing vector $\p_t$. For Boulware vacuum, its Wightman function of the scalar field is \cite{2Point-1}
\bea
G^+_B(x,x')&=&\sum_{lm}\int^\infty_0\frac{e^{-i\omega\Delta t}}{4\pi\omega}|Y_{lm}(\theta,\phi)|^2\no\\
&&~~~\times\left[|\overrightarrow{R}_l(\omega,r)|^2+|\overleftarrow{R}_l(\omega,r)|^2\right]d\omega             \label{eq3.3}
\eea
whose Fourier transformation with respect to the proper time 
\be
d \tau=\sqrt{g_{00}} d t=\sqrt{1-\frac{2 M}{r}} d t
\ee
gives 
\be
\mathcal{G}_B(\omega)
=\sum_{l}\frac{2l+1}{8\pi\omega}\left[|\overrightarrow{R}_l(\omega\sqrt{g_{00}},r)|^2+|\overleftarrow{R}_l(\omega\sqrt{g_{00}},r)|^2\right]\theta(\omega)             \label{eq3.4}
\ee
where $\theta(\omega)$ is the step function. 

Near the regions $r \rightarrow 2 M$ or $r\rightarrow \infty$, the radial functions $\overrightarrow{R_{l}}(\omega, r)$ and $\overleftarrow{R_{l}}(\omega, r)$ have asymptotic forms as summarized in \ref{appendix}. Further using the geometrical optics approximation, the Kossakowski coefficients can be calculated as
\bea
\gamma_{\pm,B}&=&\sum_{l=0}^\infty\frac{2l+1}{8\pi\omega}\left[|\overrightarrow{R}_l(\omega,r)|^2+|\overleftarrow{R}_l(\omega,r)|^2\right]\no\\
&\approx&\frac{\omega}{2\pi}\frac{1}{1-2M/r}\no\\
\gamma_B&=&\gamma_{-,B}/\gamma_{+,B}=1             \label{eq3.5}
\eea

\begin{figure}[hbtp]
\begin{center}
\subfloat[]{\includegraphics[width=.246\textwidth]{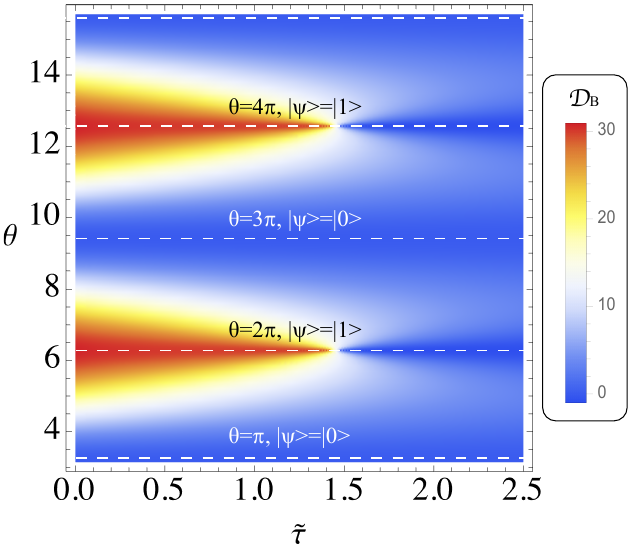}}~
\subfloat[]{\includegraphics[width=.246\textwidth]{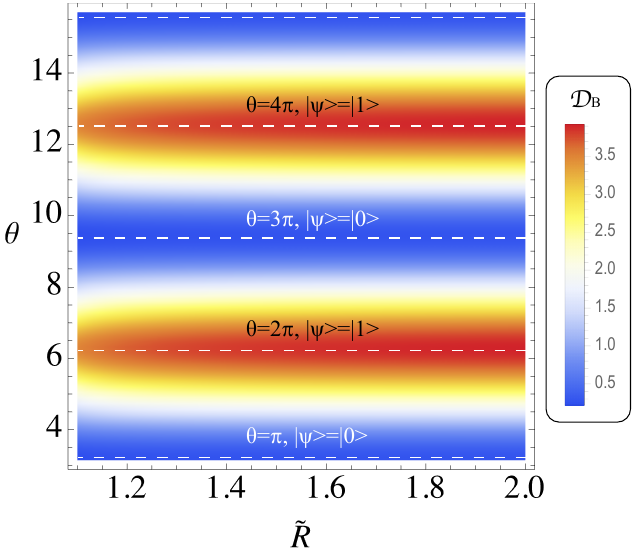}}~
\caption{The QRE of the detector interacting with background field in Boulware vacuum. (a) As a function of initial state choices $\theta$ and proper time $\tilde{\tau}$, asymptotically $\mathcal{D}_{B}=0$ indicates a unique thermal end exists. For excited initial states (white dashed lines), a sudden death of $\mathcal{D}_{B}$ happens at a critical time. The estimation is taken for $\tilde{R}=1.2$. (b) As a function of initial states $\theta$ and relative distance $\tilde{R}$ to the black hole, the detector in the ground state always has $\mathcal{D}_{B}=0$ and thus can not be excited. The estimation is taken for $\tilde{\tau}=0.2$.}
\label{fig2}
\end{center}
\end{figure}

For the detector starting from a general initial state, i.e., $|\psi\ra=\sin\frac{\theta}{2}|0\ra+\cos\frac{\theta}{2}|1\ra$, the QRE (\ref{eq3.1}) can be numerically estimated and depicted. Fig.\ref{fig2}(a) displays the thermalization process of the detector starting from arbitrary initial states. We observe that the detector evolves to a unique asymptotic state (\ref{eq2.9}) with $\mathcal{D}_{B}=0$, which is independent of the evolution paths the detector follows. In Fig.\ref{fig2}(b), we depict the QRE between the detector states and thermalization end as a function of initial state choice $ \theta$ and the relative distance to event horizon $\tilde{R}$. For general superposition initial states, we observe that the QRE at fixed time is nonvanishing and increases with larger $\tilde{R}$ from the black hole horizon. 

We note some exotic behaviors of the QRE emerging for the detector initially prepared at states with $\theta=k\pi$ ($k\in\mathbb{Z}$):

(1) For $\theta=(2k+1)\pi$, the detector starts from a ground state $|\psi\ra\sim|0\ra$. The detector Bloch vector is always parallel to the thermalized state $\bm{n}_{\text{thermal}}$, indicated by
\be
\cos\alpha_\text{ground, B}=\frac{\bm{n}(\tau)\cdot\bm{n}_{\text{thermal}}}{\ell\times\ell_{\text{thermal}}}=1
\ee
The related QRE (\ref{eq3.1}) becomes 
\be
\mathcal{D}_\text{ground, B}(\tau)=0             \label{eq3.7}
\ee 
which means the detector state cannot be distinguished from the asymptotic state (\ref{eq2.9}) all the time. However, noting that in this case the effective temperature of the Gibbs state (\ref{eq2.9}) is vanishing $T_{\text{eff}}=0$, therefore, we conclude that the detector can never be excited, consistent with our understanding of the Boulware vacuum as an empty state for large radii. In a word, in terms of QRE, we conclude that the vanishing QRE can be identified as the vanishing of the excited rate. This suggests that no thermal radiation from the black hole exists, and in the Boulware vacuum, the thermalization of a UDW detector is solely driven by its intrinsic open dynamics.

(2) For $\theta=2k\pi$, the detector starts from an excited state $|\psi\ra\sim|1\ra$. The angle between the detector state and its thermalized end is determined by the sign of the third component of (\ref{eq2.8}), which gives
\be
\cos\alpha_\text{excited, B}=\text{sign}\left(2\mathcal{E}_\text{excited, B}^2-1\right)
\ee
This means that the initially nonvanishing QRE will undergo a sudden death after a critical time
\be
\tilde{\tau}_0=4\pi\ln2\left(1-\frac{1}{\tilde{R}}\right)
\ee
just as indicated by two white dashed lines in Fig.\ref{fig2}(a).


 \subsection{The Hartle-Hawking vacuum}
The Hartle-Hawking vacuum is a state when Schwarzschild black hole is in equilibrium with a background bath which is at the same temperature of the black hole. It is constructed from Hadamard regularization and, thus, is well-behaved at the horizon. The Wightman function in the Hartle-Hawking vacuum is \cite{2Point-1,2Point-2}
\bea
G^+_H(x,x')&=&\sum_{lm}\int^\infty_0\frac{|Y_{lm}(\theta,\phi)|^2}{4\pi\omega}\left[\frac{e^{-i\omega\Delta t}}{1-e^{-2\pi\omega/\kappa}}|\overrightarrow{R}_l(\omega,r)|^2\right.\no\\
&&\left.+\frac{e^{i\omega\Delta t}}{e^{2\pi\omega/\kappa}-1}|\overleftarrow{R}_l(\omega,r)|^2\right]d\omega             \label{eq3.9}
\eea
whose Fourier transformation gives 
\bea
\mathcal{G}_H(\omega)&=&\int_{-\infty}^\infty e^{i\omega \tau} G^+_H(x,x')d\tau\no\\
&=&\sum_{l}\frac{2l+1}{8\pi\omega}\left[\frac{|\overrightarrow{R}_l(\omega\sqrt{g_{00}},r)|^2}{1-e^{-2\pi\omega\sqrt{g_{00}}/\kappa}}+\frac{|\overleftarrow{R}_l(\omega\sqrt{g_{00}},r)|^2}{1-e^{-2\pi\omega\sqrt{g_{00}}/\kappa}}\right]\no\\
             \label{eq3.10}
\eea
Close to the asymptotic regions ($r_{\text{asymp}}\rar2M$ or $r_{\text{asymp}}\rar\infty$), the Kossakowski coefficients can be given in a unified form 
\bea
\gamma_{+,H} &\approx& \frac{\omega}{2 \pi} \frac{e^{2 \pi \omega \sqrt{g_{00}}/\kappa}+1}{e^{2 \pi \omega \sqrt{g_{00}}/\kappa}-1}\left[1+g_{00} f\left(\omega \sqrt{g_{00}}, r_{\text{asymp}}\right)\right]\no \\
\gamma_{-,H} &\approx&\frac{\omega}{2 \pi}\left[1+g_{00} f\left(\omega \sqrt{g_{00}}, r_{\text{asymp}}\right)\right]           \label{eq3.11}
\eea
where the function $f(\omega,r)$ has been evaluated in (\ref{eqA4}) and (\ref{eqA5}). The ratio $\gamma=\gamma_-/\gamma_+$ becomes \cite{Open3}
\be
\gamma_H=\frac{e^{2\pi\omega\sqrt{g_{00}}/\kappa}-1}{e^{2\pi\omega\sqrt{g_{00}}/\kappa}+1}=\tanh\bigg[\frac{1}{2\tilde{T}_H}\sqrt{1-\frac{1}{\tilde{R}}}\bigg]                  \label{eq3.12}
\ee

\begin{figure}[hbtp]
\begin{center}
\subfloat[]{\includegraphics[width=.242\textwidth]{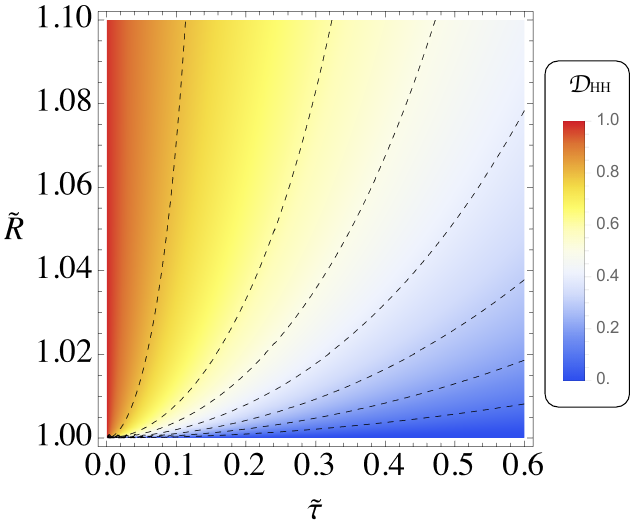}}~
\subfloat[]{\includegraphics[width=.236\textwidth]{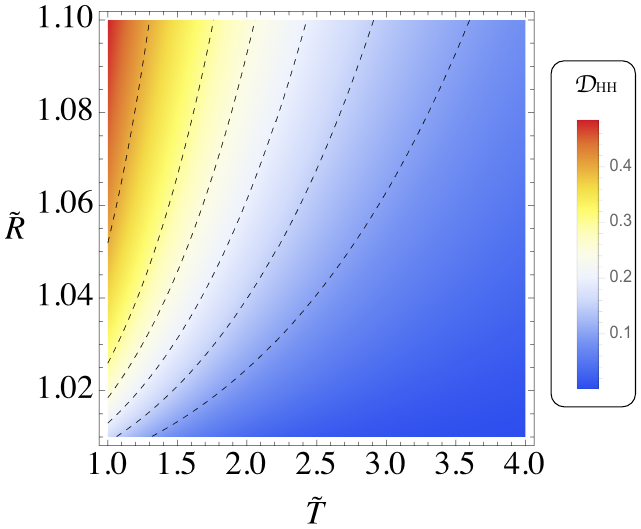}}~
\caption{The QRE of the detector in Hartle-Hawking vacuum $\mathcal{D}_{HH}$ (near horizon). (a) As a function of proper time $\tilde{\tau}$ and relative distance $\tilde{R}$ to the black hole, asymptotically $\mathcal{D}_{HH}=0$ refers that the detector is eventually thermalized to a Gibbs state with $T_{\text{eff}}$. The estimation is taken for $\tilde{T}_H=1$. (b) As a function of Hawking temperature $\tilde{T}_H$ and the relative distance $\tilde{R}$ to the black hole, degrading $\mathcal{D}_{HH}$ as approaching the horizon shows that the detector is heavily thermalized. The estimation is taken for $\tilde{\tau}=0.5$.}
\label{fig3}
\end{center}
\end{figure}

For simplicity of discussion, we consider the UDW detector initially prepared in a ground state, e.g., $\theta=\pi$. The related QRE $\mathcal{D}_{HH}$ is depicted in Fig.\ref{fig3} at the near horizon limit. 

Fig.\ref{fig3}(a) shows that the detector hovering outside the black hole eventually evolves to a unique thermalization end with $\mathcal{D}_{HH}=0$. Unlike the Boulware case, we can confirm from (\ref{eq2.9}) that this equilibrium state has a nonvanishing effective temperature
\be
T_{\text{eff}}=T_H\left(1-\frac{1}{\tilde{R}}\right)^{-1/2}       \label{eq3.14}
\ee
which approaches to the Hawking temperature $T_H$ at spatial infinity ($\gamma\rar\tanh(1/2T_H)$), but blows up at the event horizon $\tilde{R}\rar1$ ($\gamma\rar0$). This demonstrates the fact that the ground state detector in the vacuum would spontaneously excite with an excitation rate that is the same as it immersed with a flux of thermal radiation at the temperature $T_{\text{eff}}$. Moreover, we justify that the thermalization process of the detector now depends on its radial relative distance to the black hole, i.e., the detector is thermalized faster, indicating by a rapidly degrading QRE near the horizon $r_H$. For a fixed proper time, $\mathcal{D}_{HH}$ becomes a function of Hawking temperature $\tilde{T}_H$ and the relative distance $\tilde{R}$ to the black hole, shown in Fig.\ref{fig3}(b). We observe that for a black hole with larger Hawking radiation, the QRE degrades monotonously and eventually arrives at a Gibbs state with $\mathcal{D}_{HH}=0$.

Noting that the Kossakowski coefficients have the same formulas at both regions $r_{\text{asymp}}\rar2M$ and $r_{\text{asymp}}\rar\infty$, we expect that $\mathcal{D}_{HH}$ for a detector hovering at asymptotic infinity should have same behavior as it does close to the horizon.

\subsection{The Unruh vacuum}
We now move to the Unruh vacuum, which corresponds to the Boulware vacuum in the far past, and the Hartle-Hawking vacuum in the far future \cite{Hawking3}. It is considered as the best approximation of the state following the gravitational collapse of a massive body. The Wightman function for massless scalar fields in the Unruh vacuum can be computed as follows \cite{2Point-1}
\bea
G_{U}^{+}\left(x, x^{\prime}\right)&=& \sum_{m l} \int_{-\infty}^{\infty} \frac{\left|Y_{l m}(\theta, \phi)\right|^{2}}{4 \pi \omega}\left[\frac{e^{-i \omega \Delta t}}{1-e^{-2 \pi \omega / \kappa}}\left|\vec{R}_{l}(\omega, r)\right|^{2} \right.\no\\
& &\left.+\theta(\omega)e^{-i \omega \Delta t}\left|\overleftarrow{R}_{l}(\omega, r)\right|^{2}\right] d \omega      \label{eq3.15}
\eea
whose Fourier transform with respect to the proper time gives
\bea
\mathcal{G}_{U}\left(\omega\right)&=&\frac{1}{8 \pi \omega} \sum_{l=0}^{\infty}\left[\theta\left(\omega \sqrt{g_{00}}\right)(1+2 l)\left|\overleftarrow{R_{l}}\left(\omega \sqrt{g_{00}}, r\right)\right|^{2}\right.\no\\
&&+\frac{(1+2 l)\left|\overrightarrow{R_{l}}\left(\omega \sqrt{g_{00}}, r\right)\right|^{2}}{1-e^{-2 \pi \omega \sqrt{g_{00}} / \kappa}}\bigg]      \label{eq3.16}
\eea

In the asymptotic region $r_{\text{asymp}}\rar 2M$, the Kossakowski coefficients take a form 
\bea
\gamma_{+,U} &\approx& \frac{\omega}{2 \pi}\left[g_{00} f\left(\omega\sqrt{g_{00}}, r_{\text{asymp}}\right)+\frac{e^{2 \pi \omega \sqrt{g_{00}}/\kappa}+1}{e^{2 \pi \omega \sqrt{g_{00}}/\kappa}-1}\right]\no \\
\gamma_{-,U} &\approx& \frac{\omega}{2 \pi}\left[1+g_{00} f\left(\omega \sqrt{g_{00}}, r_{\text{asymp}}\right)\right]\no\\
\gamma_U&=&\frac{e^{2 \pi \omega \sqrt{g_{00}}/\kappa}-1}{e^{2 \pi \omega  \sqrt{g_{00}}/\kappa}+1}      \label{eq3.17}
\eea

\begin{figure}[hbtp]
\begin{center}
\subfloat[]{\includegraphics[width=.246\textwidth]{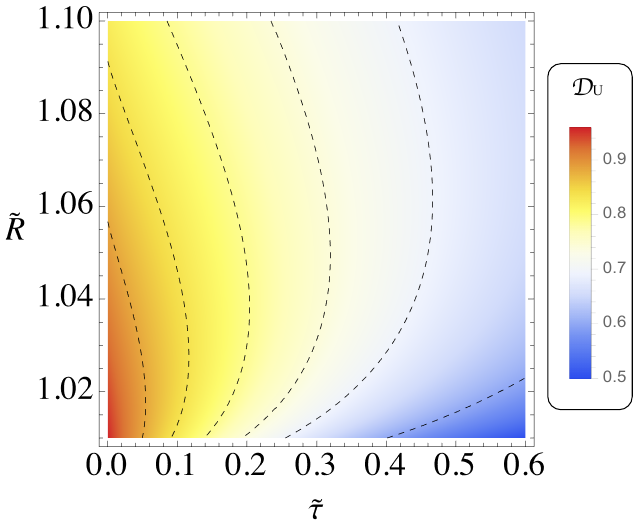}}~
\subfloat[]{\includegraphics[width=.246\textwidth]{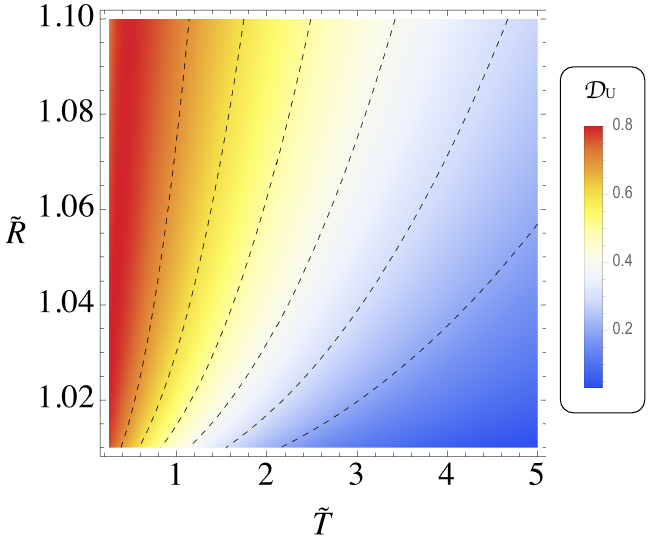}}~
\caption{The QRE of the detector interacting with background field in Unruh vacuum $\mathcal{D}_{U}$ (near horizon). (a) As a function of proper time $\tilde{\tau}$ and relative distance $\tilde{R}$, asymptotically $\mathcal{D}_{U}=0$ shows that the detector is eventually thermalized to a Gibbs state. The backscattering off the spacetime curvature may cause a more gentle thermalization for the detector sufficient away from the black hole. The estimation is taken for $\tilde{T}_H=0.3$. (b) As a function of Hawking temperature $\tilde{T}_H$ and the relative distance $\tilde{R}$, degrading $\mathcal{D}_{U}$ as approaching the horizon shows that the detector is heavily thermalized. The estimation is taken for fixed proper time $\tilde{\tau}=0.3$.}
\label{fig4}
\end{center}
\end{figure}

In virtue of (\ref{eq3.1}), it follows that the QRE of the detector $\mathcal{D}_{U}$ in the vicinity of the horizon can be numerically estimated and displayed in Fig.\ref{fig4}. 

Same as the Hartle-Hawking case, the time-evolution of QRE shown in Fig.\ref{fig4}(a) indicates that the detector approaches a unique thermalization end with $\mathcal{D}_{U}=0$ after a sufficiently long time. However, as the detector departs from the horizon, we observe the detector's thermalization encoded by the time behavior of $\mathcal{D}_{U}(\tau)$ becomes more gentle. Compared to the Hartle-Hawking case (Fig.\ref{fig3}(a)), this distinctive thermalization pattern can be attributed to a solely grey-body term $g_{00}f(\omega_0\sqrt{g_{00}},r)$ appearing in (\ref{eq3.17}), which is caused by the backscattering off the spacetime curvature and weaken the thermal flux going outwards from the horizon.

At spatial infinity $r_{\text{asymp}}\rar\infty$, the backscattering off the spacetime curvature would significantly modify the Kossakowski coefficients as
\bea
\gamma_{+,U} &\approx& \frac{\omega}{2 \pi}\left[1+g_{00} f\left(\omega\sqrt{g_{00}}, r_{\text{asymp}}\right)\frac{e^{2 \pi \omega \sqrt{g_{00}}/\kappa}+1}{e^{2 \pi \omega \sqrt{g_{00}}/\kappa}-1}\right]\no \\
\gamma_{-,U} &\approx& \frac{\omega}{2 \pi}\left[1+g_{00} f\left(\omega \sqrt{g_{00}}, r_{\text{asymp}}\right)\right]
\eea
With in mind that the grey-body factor $g_{00}f(\omega_0\sqrt{g_{00}},r)$ is vanishing asymptotically (see Eq.(\ref{eqA5}) in \ref{appendix}), we conclude that $\gamma_U\approx 1$ at spatial infinity. This is just coincident with the previous discussion of the Boulware vacuum case, thus suggesting that at spatial infinity, we always have $\mathcal{D}_{U}=0$, independent of the evolution paths the detector follows. This demonstrates the fact that, in the Unruh vacuum, no thermal radiation is perceived at spatial infinity due to the backscattering of the outgoing thermal radiation off the spacetime curvature.


\section{Quantum coherence outside a black hole}

The thermalization process of the UDW detector is an irreversible quantum process that undergoes the Hawking thermal flux from the black hole horizon. The irreversibility of the process can be well characterized by the entropy production rate satisfying quantum thermodynamic 2nd law \cite{qthermo5,qthermo4} 
\be
\Pi=-\frac{1}{T} \frac{\mathrm{d} F\left(\rho\right)}{\mathrm{d} t}\geqslant0      \label{eq4.1}
\ee
where $F(\rho)$ is the free energy of the system\footnote{Through the non-equilibrium quantum processes, the entropy change of the system is $d{S}/dt=\Pi-\Phi$, where $\Phi$ is entropy flux rate from the system to the environment, and $\Pi$ is the entropy production rate with quantum origin.}. During the thermalization process, the change of free energy can be recast into $\beta\Delta F=\mathcal{D}_{KL}+ \mathcal{C}$, an entropic-based separation. 

For an UDW detector,  we compute its Kullback-Leibler divergence $\mathcal{D}_{KL}\equiv \sum_n p_n \log {p_n}/{p_n^{\text {th }}}$, where $p_n=\left\langle n\left|\rho\right| n\right\rangle$ are the population of a density matrix, as
\be
\mathcal{D}_{KL}=2\log 2+\log \frac{1-n^2_3}{1-\gamma^2}+n_3\log \frac{(1+n_3)(1+\gamma)}{(1-n_3)(1-\gamma)}          \label{eq4.2}
\ee
It quantifies the free energy increment due to population imbalance with respect to the thermalization end configuration and is, therefore, a purely classical term. 

The quantum coherence term $\mathcal{C}$, which is of a genuine quantum nature, determines the surplus in free energy that a non-equilibrium state with quantum coherences offers with respect to its diagonal (and thus classical) counterpart. {It is natural then to quantify the coherence of a general state $\rho$ with respect to some incoherent states, those with diagonal density matrix. As was proven in \cite{Coherence1}, the most satisfied distance-based quantifier is a particular QRE, which selects out the minimal distance between target state $\rho$ to an incoherent state set $\mathcal{I}$
\be
\mathcal{C}(\rho)=\min _{\sigma \in \mathcal{I}} S(\rho \| \sigma)
\ee
Defining the dephasing operation on density matrix as ${\rho}_{\text {diag}}=\sum_{i} \rho_{i i}\left| i\right\rangle\left\langle i\right|$, and straightforwardly using the relation $\operatorname{Tr}\left[\rho \log  \sigma\right]=\operatorname{Tr}[{\rho}_{\text {diag}} \log  \sigma]$, one can prove that 
\be
S(\rho \| \sigma)=S({\rho}_{\text {diag }})-S(\rho)+S({\rho}_{\text {diag }} \| \sigma)
\ee
Minimizing the right-hand side of the above equation over all incoherent states $\sigma \in \mathcal{I}$, we immediately see that the minimum is achieved for $\sigma={\rho}_{\text {diag}}$, which is just QRE (\ref{eq1.4}) and can be recast to
\be
\mathcal{C}({\rho})=S\left({\rho}_{\text {diag }}\right)-S({\rho})
\ee}

For an UDW detector with dynamics (\ref{eq2.8}), the QRE of quantum coherence is \cite{Open22}
\be
\mathcal{C}(\tau)=H_{\text{bin}}\left(\frac{1+n_3(\tau)}{2}\right)-H_{\text{bin}}\left(\frac{1+\ell}{2}\right)     \label{eq4.3}
\ee
where $H_{\text {bin}}(x) \equiv-x \log x-(1-x) \log (1-x)$ is a binary entropy of variable $x$. It is a time-dependent function of the detector's initial state preparation $\theta$, the rescaling Hawking temperature $\tilde{T}_H$, and the relative distance $\tilde{R}$ of the detector to the horizon. 

Without loss of generality, hereafter, we work in a particular detector initial state with $\theta=\pi/2$, having a nonvanishing initial coherence. Since the nontrivial influence from Hawking radiation is the most concerning, we thus concentrate on the nonequilibrium open process of the detector with respect to the Hartle-Hawking vacuum.

\begin{figure}[hbtp]
\begin{center}
{\includegraphics[width=.45\textwidth]{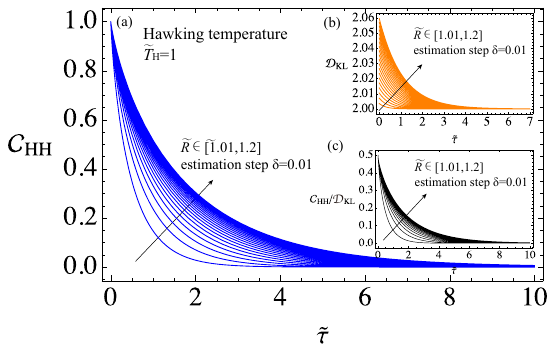}}
\caption{{The time evolution of (a) quantum coherence and (b) classical Kullback-Leibler divergence for a thermalization process in Hartle-Hawking vacuum for different values of relative distance $\tilde{R}$ (near horizon). The ratio between these two contributions in the thermalization process as a function of the detector's proper time is depicted in (c). } }
\label{fig5}
\end{center}
\end{figure}

In Fig.\ref{fig5}, we depict the quantum coherence monotone $\mathcal{C}_{HH}$ and Kullback-Leibler divergence $\mathcal{D}_{KL}$ for the thermalization process of a UDW detector with respect to Hartle-Hawking vacuum. {As proper time passed, it is obvious that both quantum coherence and classical Kullback-Leibler divergence are consumed during the open process until the detector approaches its thermalization end. As the relative distance $R$ grows, we observe a slower decrement of quantum coherence. This could be attributed to the previous observation that detector thermalization becomes more gently as it departs away from the horizon and eventually arrives at the (diagonal) thermal end with a lower effective temperature.}  

\begin{figure}[hbtp]
\begin{center}
\subfloat[]{\includegraphics[width=.23\textwidth]{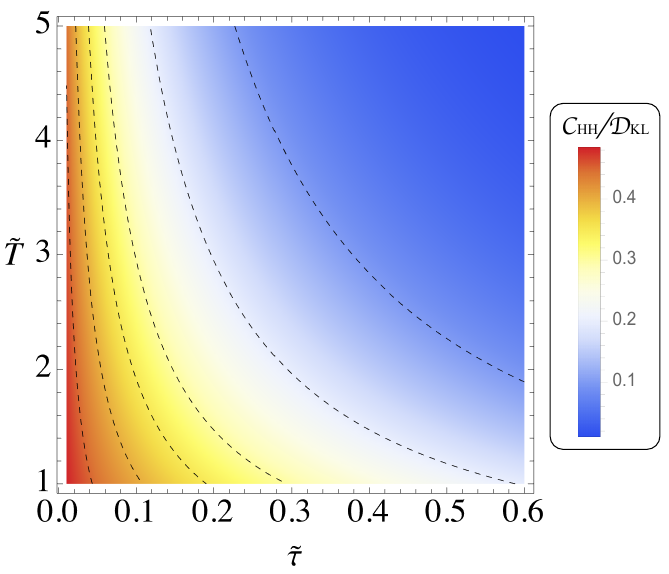}}~
\subfloat[]{\includegraphics[width=.25\textwidth]{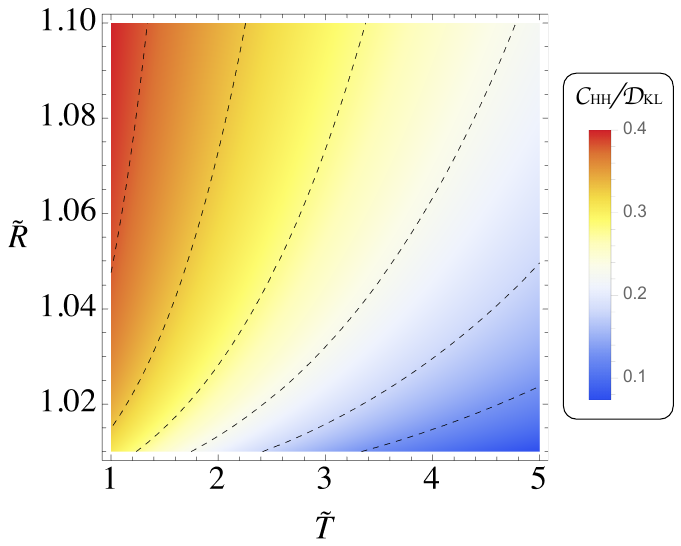}}~
\caption{The ratio $\mathcal{C}_{HH}/\mathcal{D}_{KL}$ between quantum (coherence) and classical (Kullback-Leibler) contributions in thermalization process with respect to the Hartle-Hawking vacuum (near horizon). (a) The ratio as a function of proper time $\tilde{\tau}$ and Hawking temperature $\tilde{T}_H$, showing different consumption rates of quantum and classical resources during the thermalization process. The estimation is taken for $\tilde{R}=1.02$. (b) The ratio is a function of Hawking temperature $\tilde{T}_H$ and the relative distance $\tilde{R}$ to the black hole. The estimation is taken for $\tilde{\tau}=0.2$.}
\label{fig6}
\end{center}
\end{figure}

It is interesting to note that during the thermalization process, the consumption \emph{rate} of quantum coherence is quite different from that of its classical counterpart. {In Fig.\ref{fig5}(c), we illustrate that for the detector hovering at a specific relative distance, how the ratio between quantum coherence and classical Kullback-Leibler divergence, i.e.,  $\mathcal{C}_{HH}/\mathcal{D}_{KL}$, changes with respect to detector's proper time. One can see that the ratio degrades monotonously over time, which means that the quantum coherence of the detector is consumed faster than its Kullback-Leibler divergence during the open process. From a quantum thermodynamic perspective (see Eq.(\ref{eq1.3}) and (\ref{eq4.1})), we recognize that a significant contribution to the entropy production comes from the quantum coherence consumption.}

{To further distinguish the coherence and Kullback-Leibler contributions during the thermalization process, the influence of Hawking effect on the ratio $\mathcal{C}_{HH}/\mathcal{D}_{KL}$ is displayed in Fig.6. In particular, we observe that as Hawking temperature growing, the ratio function decays faster which indicates a larger consumption rate of quantum coherence than to Kullback-Leibler divergence} is required in a thermalization process. In Fig.\ref{fig6}(b), the ratio $\mathcal{C}_{HH}/\mathcal{D}_{KL}$ is illustrated at a fixed time as a function of Hawking temperature and detector's relative distance to the black hole. We see a nearly vanishing ratio as approaching the event horizon, meaning that the quantum coherence is so drastically consumed compared to its classical counterpart, consistent with the fact the effective temperature is blowing up at the event horizon. When departing away from the black hole, the growing ratio means less consumption of quantum resources for the detector's thermalizing to a Gibbs end.


\section{Conclusions}
\label{4}

In this work, we employ the quantum relative entropy to revise the thermalization process of a UDW detector outside a Schwarzschild black hole. Beyond detailed balance conditions or Planckian transition rates, the QRE between the detector density matrix and its thermalization end is a feature function of particular thermalization {process}. It was shown that a ground state detector interacting with the background field in a Boulware vacuum has vanishing QRE all the time, which means that the detector can not be excited. With respect to the Hartle-Hawking vacuum, the detector QRE degrades monotonously for larger Hawking radiation or close to the event horizon, attributed to the fact that the effective Hawking temperature depends on the radial position of the detector and blows up at the event horizon. With respect to an Unruh vacuum, however, the detector QRE becomes non-monotonous as the detector departs away from the horizon, indicating a more gentle time behavior of the detector thermalization. 

From a perspective of quantum thermodynamics, we evaluate a particular form of QRE as a measure of quantum coherence, a genuine quantum contribution driving the thermodynamic irreversibility of the thermalization process. For the Hartle-Hawking vacuum, we find that during the open process, {the QRE of quantum coherence monotonously degrades over time, similar to the case of accelerating UDW detector in Minkowski spacetime \cite{Open20,Open21,Open22}. Nevertheless, the position-dependent effective Hawking temperature (\ref{eq3.14}) results in additional dependence of detector coherence on the relative distance to the event horizon. On the other hand, beyond other feature functions of the thermalization process (such as $l_1-$norm \cite{Open20,Open21} or quantum Fisher information \cite{Open22}), the QRE coherence monotone enables us to refine the thermalization process by decomposing the entropy production into genuine quantum and classical contributions}. In particular, we find that the consumption rate of quantum coherence could be quite different from that of its classical counterpart. For example, as Hawking temperature grows, a larger consumption rate of quantum coherence than classical Kullback-Leibler divergence is needed in a thermalization process. {}

We believe that our analysis of the entropic formulation of the thermalization may have broad interests, as many quantum effects in gravity can be reinterpreted in a similar way. In fact, some research on quantum thermal engines utilizing the Unruh effect has been proposed \cite{qthermo6,qthermo7,qthermo8}. It may deepen our interpretation of quantum gravity if such exploration in black hole spacetimes is made in the future \cite{qthermo9}.

\section*{Acknowledgement}
This work is supported by the National Natural Science Foundation of China (No.12475061, 12075178) and the Shaanxi Fundamental Science Research Project for Mathematics and Physics (No. 23JSY006). S.H. and Z.O. contributed equally to this work. J. F. thanks Liu Zhao and Wen-Jing Chen for the stimulating discussions.
\appendix

\section{}

\label{appendix}
We summarize here a number of formulas that are useful in the evaluation of the results in Section \ref{sec3}. More details can be found in Ref.\cite{2Point-1,2Point-2}. 

Throughout the discussion, we are interested in the QRE in the two asymptotic regimes $r \rightarrow 2 M$ and $r \rightarrow \infty$, where the radial functions $\overrightarrow{R_{l}}(\omega, r)$ and $\overleftarrow{R_{l}}(\omega, r)$ have asymptotic forms like
\be
\sum_{l=0}^{\infty}(2 l+1)|\overrightarrow{{R}}_{l}(\omega \mid r)|^{2} 
\sim\left\{\begin{array}{l}
\displaystyle 4 \omega^{2}\left(1-2 M / r\right)^{-1}, ~~r \rightarrow 2 M \\
\displaystyle\frac{1}{r^{2}} \sum_{l=0}^{\infty}(2 l+1)\left|B_{l}(\omega)\right|^{2}, ~~r \rightarrow \infty
\end{array}\right.
\ee

\be
\sum_{l=0}^{\infty}(2 l+1)|\overleftarrow{{R}}_{l}(\omega \mid r)|^{2} 
\sim\left\{\begin{array}{l}
\displaystyle\frac{1}{4 M^{2}} \sum_{l=0}^{\infty}(2 l+1)\left|B_{l}(\omega)\right|^{2},~~ r \rightarrow 2 M \\
\displaystyle4 \omega^{2}, ~~ r \rightarrow \infty
\end{array}\right.
\ee

With a sustained cut-off point around $\sqrt{27} M \omega$, the transmission amplitude $B_{l}(\omega)$ can be evaluated through the geometrical optics approximation, which gives
\be
B_{l}(\omega) \sim \theta(\sqrt{27} M \omega-l)
\ee
This enables us to evaluate the function $f(\omega,r)$ in (\ref{eq3.11}) as
\be
f(\omega,r)\equiv \frac{1}{4r^2\omega^2}\sum_{l=0}^\infty(2l+1)\left|B_{l}(\omega)\right|^2                \label{eqA4}
\ee
which leads to the combination in terms of relative distance (\ref{eq3.8})
\be
g_{00}f(\omega_0\sqrt{g_{00}},r)\sim\frac{27}{16\tilde{R}^2}\left(1-\frac{1}{\tilde{R}}\right)               \label{eqA5}
\ee
approaching zero in both asymptotic regions.

\end{document}